\documentclass[12pt]{article}

\begin{document}

\title{\bf Two-Loop Superstrings in Hyperelliptic Language III:
the Four-Particle Amplitude}

\author{Zhu-Jun Zheng\thanks{Supported in part by Math. Tianyuan Fund
with grant Number 10226002 and the Natural Science Foundation of
Educational Committee of Henan Province with grant Number
2000110010
 } \\ \\
Institute of Mathematics, Henan University \\
Kaifeng 475001, P. R. China\\ and\\
Institute of Theoretical Physics,
Chinese Academy of Sciences\\
P. O. Box 2735,  Beijing 100080, P. R. China \\  \\
Jun-Bao Wu \\School of Physics, Peking University \\
Beijing 100871, P. R. China\\ \\
Chuan-Jie Zhu\thanks{Supported in
part by fund from the National Natural Science Foundation of China
with grant Number
90103004.} \\
Institute of Theoretical Physics,
Chinese Academy of Sciences\\
P. O. Box 2735,  Beijing 100080, P. R. China}

\maketitle

\newpage

\begin{abstract}
We compute explicitly the four-particle amplitude in superstring
theories by using the hyperelliptic language and the newly
obtained chiral measure of D'Hoker and Phong. Although the algebra
of the intermediate steps is a little bit involved, we obtain a
quite simple expression for the four-particle amplitude.  As
expected, the integrand is independent of all the insertion
points. As an application of the obtained result, we show that the
perturbative correction to the $R^4$ term in type II superstring
theories is vanishing point-wise in (even) moduli space at two
loops.
\end{abstract}


\section{Introduction}

Although we believe that superstring theory is finite in
perturbation at any order \cite{GreenSchwarz1, GreenSchwarz2,
GreenSchwarz3,Martinec}, a rigorous proof is still lacking despite
great advances in the covariant formulation of superstring
perturbation theory \'a la Polyakov. The main problem is the
presence of supermoduli and modular invariance in higher genus. At
two loops these problems were solved explicitly by using the
hyperelliptic formalism in a series of papers
\cite{GavaIengoSotkov, IengoZhu1, Zhu, IengoZhu2, IengoZhu3}. The
explicit result was also used by Iengo \cite{Iengo} to prove the
vanishing of perturbative correction to the $R^4$ term
\cite{GrossWitten} at two loop, in agreement with the indirect
argument of Green and Gutperle \cite{GreenGutperle}, Green,
Gutperle and Vanhove \cite{Green2}, and Green and Sethi
\cite{GreenSethi} that the $R^4$ term does not receive
perturbative contributions beyond one loop. Recently, Stieberger
and Taylor \cite{Stieberger}  also used the result of
\cite{IengoZhu2} to prove the vanishing of the heterotic two-loop
$F^4$ term. For some closely related works we refer the reader to
the reviews \cite{Green3, Kiritsis}. In the general case in
superstring perturbation theory, there is no satisfactory
solution. For a review of these problem we refer the reader to
\cite{DHokerPhong1, DHokerPhong6}.

Recently two-loop superstring was studied by D'Hoker and Phong. In
a series of papers \cite{DHokerPhong2, DHokerPhong3, DHokerPhong4,
DHokerPhong5} (for a recent review see \cite{DHokerPhong6}),
D'Hoker and Phong found an unambiguous and slice-independent
two-loop superstring measure on moduli space for even spin
structure from first principles.

Although their result is quite explicit, it is still a difficult
problem to use it in actual computation. In \cite{DHokerPhong5},
D'Hoker and Phong used their result to compute explicitly the
chiral measure by choosing the split gauge and proved the
vanishing of the cosmological constant and the non-renormalization
theorem \cite{DHokerPhong7, Martinec}. They also computed the
four-particle amplitude in another forthcoming paper
\cite{DHokerPhong8}. Although the final results are exactly the
expected, their computation is quite difficult to follow because
of the use of theta functions.\footnote{In \cite{Lechtenfeld5},
the two-loop 4-particle amplitude was also computed by using theta
functions. Its relation with the previous explicit result
\cite{IengoZhu2} has not been clarified.}  Also modular invariance
is absurd in their computations because of the complicated
dependence between the 2 insertion points (the insertion points
are also spin structure dependent).

In the old works \cite{GavaIengoSotkov, IengoZhu1, Zhu, IengoZhu2}
on two-loop superstrings, one of the author (with Iengo) used the
hyperelliptic representation to do the explicit computation at two
loops which is quite explicit and modular invariance is manifest
at every stage of the computations. So it is natural to do
computations in this language by using the newly established
result. In a previous paper \cite{AllZhu1}, we reported the main
results of our computation of two loop superstring theory by using
hyperelliptic language.  In \cite{AllZhu2} the details for the
verification of the vanishing of the cosmological constant and the
non-renormalization theorem is given. As we have shown in
\cite{AllZhu2}, except one kind term involving the stress energy
tensor of the $\psi$ field, all the rest relevant terms containing
3 or less particles are vanishing point-wise in moduli space after
summation over spin structures. This non-vanishing term also gives
a vanishing contribution to the 3-particle amplitude because of
the antisymmetric property of the relevant kinematic factor. As we
will see in this paper, the non-renormalization theorem greatly
simplifies the computation of the possibly non-vanishing
four-particle amplitude.

In this paper we will present the details for the explicit
computation of the four-particle amplitude. As announced in
\cite{AllZhu1}, we obtained a quite simple  and explicit
expression for the chiral integrand. For type II superstring
theories, we also give an explicit expression for the amplitude.
By using this result, we   show that the perturbative correction
to the $R^4$ term in type II superstring theories is vanishing at
two loops, confirming the computation of Iengo \cite{Iengo} and
the the conjecture of Green and Gutperle \cite{GreenGutperle} (see
also \cite{Green2, GreenSethi}). We leave the proof of the
equivalence between the new result and the old result of the
four-particle amplitude as a problem of the future. This paper is
organized as follows:

In the next section we will recall the relevant results of
hyperelliptic representation of the genus 2 Riemann surface and
set our notations for all the correlators. In section 3 we recall
the results of D'Hoker and Phong for the chiral measure. In
section 4 we computed explicitly all the relevant quantities
appearing in the chiral measure and write the chiral integrand
(before the summation over spin structures) by using hyperelliptic
language. In section 5 we present the computation of the connected
terms from the matter supercurrent and show that it gives
vanishing contribution to the amplitude. In section 6, we will use
the result of \cite{AllZhu2} and computed the terms with a single
$S_n(q)$ factor. We show that the resulting expression is
independent of the three arbitrary points $p_{1,2,3}$. In section
7 we compute the rest terms and found an almost total cancellation
between the various terms. In section 8, we present the final form
for the 4-particle amplitude by combing all the previous results.
The final result for the 4-particle amplitude is quite simple. We
also show that the result is independent of the insertion points
$q_{1,2}$. By using this result, we  show that the perturbative
correction to the $R^4$ term in type II superstring theories is
vanishing at two loops, confirming the explicit computation of
Iengo \cite{Iengo} and the conjecture of Green and Gutperle
\cite{GreenGutperle}. Some technical parts are relegated to the
appendix.  In Appendix A, we collect all the formulas for the
$\langle X(z) X(w) \rangle$ correlators. In Appendix B, we present
the proof for the summation formulas eq.~(\ref{eq999}). In
Appendix C, we compute explicitly the $q$ dependence of the factor
${\cal Z}$. Here we also give some useful formulas which are
needed in the transform from the branch point parametrization of
the moduli space to the period matrix $\tau_{ij}$.

To conclude this section we note that D'Hoker and Phong have also
proved that the cosmological constant and the 1-, 2- and 3-point
functions are zero point-wise in moduli space \cite{DHokerPhong7}.
They have also computed the 4-particle amplitude
\cite{DHokerPhong8}. The agreement of the results from these two
different gauge choices and two different methods of computations
would be the final proof of the validity of the new supersymmetric
gauge fixing method at two loops.

\section{Genus 2 hyperelliptic Riemann surface}

First we remind that a genus-g Riemann surface, which is the
appropriate world sheet for one and two loops, can be described in
full generality by means of the hyperelliptic
formalism.\footnote{Some early works on two loops computation by
using hyperelliptic representation are \cite{Knizhnik, Morozov,
Morozov1, Morozov2, Lechtenfeld, Bershadsky, Moore} which is by no
means the complete list.} This is based on a representation of the
surface as two sheet covering of the complex plane described by
the equation:
\begin{equation}
y^2(z) = \prod_{i=1}^{2g+2} ( z - a_i), \label{covering}
\end{equation}
The complex numbers $a_{i}$, $(i=1,\cdots,2g+2)$ are the $2g+2$
branch points, by going around them one passes from one sheet to
the other. For two-loop ($g=2$) three of them represent the moduli
of the genus 2 Riemann surface over which the integration is
performed, while the other three can be arbitrarily fixed. Another
parametrization of the moduli space is given by the period matrix
$\tau$. The transformation from the period matrix $\tau$ to the
branch points can be found in Appendix C.

There are two independent holomorphic differentials on a genus two
Riemann surface. In hyperelliptic language these are given
explicitly as follows:
\begin{equation}
\Omega_1(z) = { {\rm d} z \over y(z) } , \qquad \Omega_2(z) = {
z\, {\rm d} z \over y(z) } .
\end{equation}
These differentials are not normalized in the standard way:
\begin{equation}
\oint_{\alpha_i} \omega_j(z) = \delta_{ij}, \qquad \oint_{\beta_i}
\omega_j(z) = \tau_{ij}, \qquad i,j =1,2,
\end{equation}
where $\alpha_i$ and $\beta_i$ are the four nontrivial one cycles
and $\tau_{ij}$ is the $2\times2$ period matrix. Setting
\begin{equation}
(K)_{ij} = \oint_{\alpha_i} \Omega_j,
\end{equation}
we have
\begin{equation} \Omega_i = \omega_j K_{ji},
\end{equation}
and
\begin{eqnarray}
\omega_1 & = & { K_{22}\Omega_1 - K_{21} \Omega_2 \over {\rm
det}\, K } , \\
\omega_2 & = & { -K_{12}\Omega_1 + K_{11} \Omega_2 \over {\rm
det}\, K } .
\end{eqnarray}

At genus 2, by choosing a canonical homology basis of cycles we
have the following list of 10 even spin structures:
\begin{eqnarray}
\delta_1 \sim \left[ \begin{array}{cc} 1 & 1\\ 1 & 1 \end{array}
\right]  \sim (a_1 a_2  a_3|a_4 a_5 a_6),  & & \delta_2 \sim
\left[
\begin{array}{cc} 1 & 1\\ 0 & 0 \end{array} \right] \sim
(a_1a_2a_4|a_3a_5a_6), \nonumber
\\
\delta_3 \sim \left[ \begin{array}{cc} 1 & 0\\ 0 & 0 \end{array}
\right]  \sim (a_1a_2a_5|a_3a_4a_6),  & & \delta_4 \sim \left[
\begin{array}{cc} 1 & 0\\ 0 & 1 \end{array} \right] \sim
(a_1 a_2 a_6|a_3 a_4 a_5), \nonumber
\\
\delta_5 \sim \left[ \begin{array}{cc} 0 & 1\\ 0 & 0 \end{array}
\right]  \sim (a_1 a_3 a_4|a_2 a_5 a_6), & & \delta_6 \sim \left[
\begin{array}{cc} 0 & 0\\ 0 & 0 \end{array} \right] \sim
(a_1 a_3 a_5|a_2 a_4 a_6), \nonumber
\\
\delta_7 \sim \left[ \begin{array}{cc} 0 & 0\\ 0 & 1 \end{array}
\right]  \sim (a_1 a_3 a_6|a_2 a_4 a_5), & & \delta_8 \sim \left[
\begin{array}{cc} 0 & 0\\ 1 & 1 \end{array} \right] \sim
(a_1 a_4 a_5|a_2 a_3 a_6), \nonumber
\\
\delta_9 \sim \left[ \begin{array}{cc} 0 & 0\\ 1 & 0 \end{array}
\right]  \sim (a_1 a_4 a_6|a_2 a_3 a_5), & & \delta_{10} \sim
\left[
\begin{array}{cc} 0 & 1\\ 1 & 0 \end{array} \right] \sim
(a_1 a_5 a_6|a_2 a_3 a_4). \nonumber
\end{eqnarray}
We will denote an even spin structure as $(A_1 A_2 A_3|B_1 B_2
B_3)$. By convention $A_1= a_1$. For each even spin structure we
have a spin structure dependent factor from determinants which is
given as follows \cite{GavaIengoSotkov}:
\begin{equation}
Q_\delta = \prod_{i <j} (A_i-A_j)(B_i-B_j).
\end{equation}
This is a degree 6 homogeneous polynomials in $a_i$.

At two loops there are two odd supermoduli and this gives two
insertions of supercurrent  at two different points $x_1$ and
$x_2$.  Previously the chiral measure was derived in
\cite{Verlinde,DHokerPhong1} by a simple projection from the
supermoduli space to the even moduli space. This projection does't
preserve supersymmetry and there is a residual dependence on the
two insertion points. This formalism was used in
\cite{GavaIengoSotkov, IengoZhu1, Zhu, IengoZhu2}. In these papers
we found that it is quite convenient to choose these two insertion
points as the two zeroes of a holomorphic abelian differential
which are moduli independent points on the Riemann surface. In
hyperelliptic language the holomorphic abelian differential can be
written generally as follows:
\begin{equation}
\omega_x(z) = { c (z-x) \, {\rm d} z \over y(z) } .
\end{equation}
In hyperelliptic language the  two zero points are the same points
on the upper and lower sheet of the surface. We denote these two
points as $x_1=x+$ (on the upper sheet) and $x_2=x-$ (on the lower
sheet).

In the following we will give some formulas in hyperelliptic
representation which will be used later. First all the relevant
correlators are given by\footnote{We follow closely the notation
of \cite{DHokerPhong3}. }
\begin{eqnarray}
\langle \psi^\mu   (z) \psi^\nu   (w) \rangle & = &
-\delta^{\mu\nu} G_{1/2}[\delta] (z,w) = - \delta^{\mu\nu}S_\delta
(z,w),
\nonumber \\
\langle b(z) c(w) \rangle &=& +G_2 (z,w), \nonumber \\
\langle \beta (z) \gamma (w) \rangle &=& -G_{3/2}[\delta] (z,w),
\end{eqnarray}
where
\begin{eqnarray}
& & S_{\delta}(z,w) = { 1\over z-w} \, { u(z) + u(w) \over 2
\sqrt{u(z) u(w) } } , \label{eqszego} \\
& & u(z) = \prod_{i=1}^3 \left( z-A_i \over z-B_i\right)^{1/2}, \\
& & G_2(z,w) = -H(w,z) + \sum_{a=1}^3 H(w,p_a) \, \varpi_a(z,z),
 \label{eq6} \\
& & H(w,z) = { 1\over 2(w- z)}
\,\left( 1 + { y(w) \over
y(z) }\right) \, { y(w) \over y(z) }, \\
& & G_{3/2}[\delta](z,w) = - P(w,z) + P(w,q_1) \psi_1^*(z) +
P(w,q_2)\psi_2^*(z), \label{eq51} \\
& & P(w,z) = {1\over \Omega(w)}\, S_{\delta}(w,z)\Omega(z),
\end{eqnarray}
where $\Omega(z)$ is a holomorphic abelian differential satisfying
$\Omega(q_{1,2}) \neq 0$ and otherwise arbitrary. These
correlators were adapted from \cite{Iengo2}. $\varpi_a(z,w)$ are
defined in \cite{DHokerPhong2} and $\psi^*_{1,2}(z)$ are the two
holomorphic $3\over 2$-differentials. When no confusion is
possible, the dependence on the spin structure $[\delta]$ will not
be exhibited. The formulas for the $\langle X(z) X(w) \rangle$ and
related correlators are given in Appendix A.

In order take the limit of $x_{1,2}\to q_{1,2}$ we need the
following expansions:
\begin{eqnarray}
G_{3/2} (x_2, x_1) &=& {1 \over x_1 - q_1} \psi ^* _1 (x_2)
      - \psi ^* _1 (x_2) f_{3/2} ^{(1)} (x_2) +O(x_1 - q_1),
\\
G_{3/2} (x_1, x_2) &=& {1 \over x_2 - q_2} \psi ^* _2 (x_1)
      - \psi ^* _2 (x_1) f_{3/2} ^{(2)} (x_1)  +O(x_2 - q_2),
\end{eqnarray}
for $x_{1,2}  \to q_{1,2}$. By using the explicit expression of
$G_{3/2}$ in (\ref{eq51}) we have
\begin{eqnarray}
f_{3/2} ^{(1)} (q_2) & = & - {\partial_{q_2} S(q_1,q_2) \over
S(q_1,q_2)
} + \partial\psi^*_2(q_2), \label{eq54}\\
f_{3/2} ^{(2)} (q_1) & = &   {\partial_{q_1} S(q_2,q_1) \over
S(q_1,q_2) } + \partial\psi^*_1(q_1) = f_{3/2} ^{(1)}(q_2)|_{ q_1
\leftrightarrow q_2 } . \label{eq55}
\end{eqnarray}

The quantity $\psi^*_\alpha (z)$'s are holomorphic $3\over
2$-differentials and are constructed as follows:
\begin{equation}
\psi^*_\alpha (z) = (z-q_\alpha)S(z,q_\alpha)  \, {y(q_\alpha
)\over y(z)} \, , \qquad \alpha = 1, 2.
\end{equation}
For $z=q_{1,2}$ we have
\begin{eqnarray} &  & \psi^*_\alpha (q_\beta ) =
\delta_{\alpha \beta}, \label{psinormal} \\
& & \partial \psi^*_1 (q_2) = -\partial \psi^*_2 (q_1) =
S(q_1,q_2) = {i\over 4}S_1(q), \\
& & \partial \psi^*_1 (q_1) =  \partial \psi^*_2 (q_2) =
- {1\over 2} \Delta_1(q),  \\
& & \partial^2  \psi^*_1 (q_1) =  \partial^2 \psi^*_2 (q_2) =
{1\over 16}S_1^2(q)  + {1\over 4}\Delta_1^2(q) + {1\over
2}\Delta_2(q),
\end{eqnarray}
where
\begin{eqnarray}
\Delta_n(x) & \equiv & \sum_{i=1}^6 {
1\over (x - a_i)^n }, \\
S_n(x) & \equiv &   \sum_{i=1}^3 \left[ { 1\over (x - A_i)^n } - {
1\over (x - B_i)^n }\right],
\end{eqnarray}
for $  n = 1, 2$. This shows that $\partial\psi^*_\alpha
(q_{\alpha+1})$ and $\partial^2\psi^*_\alpha(q_{\alpha})$ are spin
structure dependent.

The other quantities introduced in \cite{DHokerPhong1} which will
be used later are as follows:
\begin{eqnarray}
\varpi_1(q_1,q_2) & = & - { y^2( p_1)\over y^2(q) }\, { (q-
p_2)(q-  p_3) \over  (p_1-  p_2)(p_1- p_3) } \, ,
\\
\varpi_2(q_1,q_2) & = & - { y^2(p_2)\over y^2(q) }\, {
(q-p_1)(q-p_3) \over  (p_2-p_1)(p_2-p_3) } \, ,
\\
\varpi_3(q_1,q_2) & = & - { y^2(p_3)\over y^2(q) }\, {
(q-p_1)(q-p_2) \over  (p_3-p_1)(p_3-p_2) } \, ,
\end{eqnarray}
and
\begin{eqnarray}
\varpi^*_1(u) & = & { y ( p_1)\over y (u) }\, { (u p_1 - {1\over
2}(u+p_1)(p_2+p_3) + p_2p_3)\over (p_1-p_2)(p_1-p_3)} \nonumber
\\
& = & { y ( p_1)\over y (u) } \left[ 1 + { 1\over 2} \, (u-p_1)
\,\left( { 1\over p_1-p_2 }
 + { 1\over p_1-p_3} \right) \right] \, ,
\\
\varpi^*_2(u) & = & { y ( p_2)\over y (u) }\, { (u p_2 - {1\over
2}(u+p_2)(p_3+p_1) + p_1p_3)\over (p_2-p_3)(p_2-p_1)} \, ,
\\
\varpi^*_3(u) & = & { y ( p_3)\over y (u) }\, { (u p_3 - {1\over
2}(u+p_3)(p_1+p_2) + p_1p_2)\over (p_3-p_1)(p_3-p_2)} \, .
\end{eqnarray}
We note here that $ \varpi_1(q_1,q_2) = -1$ and $
\varpi_{2,3}(q_1,q_2) = 0$ in the limit $ p_1 \to q_{1,2}$.

\section{The chiral measure: the result of D'Hoker and Phong}

In this section we will briefly recall the result of D'Hoker and
Phong.  Previously the chiral measure was derived in
\cite{Verlinde, DHokerPhong1} by a simple projection from the
supermoduli space to the even moduli space. This projection does't
preserve supersymmetry and there is a residual dependence on the
two insertion points. In the new formalism of D'Hoker and Phong,
they use a  supersymmetric projection to the super-period matrix.
For detailed derivation of the chiral measure we refer the reader
to their original papers \cite{DHokerPhong2, DHokerPhong3,
DHokerPhong4, DHokerPhong5}. After making the choice $x_\alpha =
q_\alpha$ ($\alpha= 1, 2$), the chiral measure obtained in these
papers is:
\begin{eqnarray}
{\cal A} [\delta] & = & i {\cal Z} \biggl \{ 1  + {\cal X}_1 + {\cal
X}_2 + {\cal X}_3 + {\cal X}_4 +  {\cal X}_5 + {\cal X}_6 \biggr
\},
\nonumber \\
{\cal Z} & = & {\langle  \prod _a b(p_a) \prod _\alpha \delta
(\beta (q_\alpha)) \rangle \over \det \omega _I \omega _J (p_a) }
, \label{eqcalx}
\end{eqnarray}
and the ${\cal X}_i$ are given by:
\begin{eqnarray}
{\cal X}_1 + {\cal X}_6 &=& {\zeta ^1 \zeta ^2 \over 16 \pi ^2}
\biggl [ -\langle \psi(q_1)\cdot \partial X(q_1) \, \psi(q_2)\cdot
\partial X(q_2) \rangle  \nonumber  \\
&& \hskip -1cm
 - \partial_{q_1} G_2 (q_1,q_2) \partial\psi^*_1 (q_2)
 + \partial_{q_2} G_2 (q_2,q_1) \partial\psi^*_2 (q_1)
\nonumber \\
&& \hskip -1cm + 2   G_2 (q_1,q_2) \partial\psi^*_1 (q_2)  f_{3/2}
^{(1)} (q_2) - 2   G_2 (q_2,q_1) \partial\psi^*_2 (q_1)  f_{3/2}
^{(2)} (q_1) \biggr ] \, ,
 \\
{\cal X}_2 + {\cal X}_3 &=&  {\zeta ^1 \zeta ^2 \over 8 \pi ^2}
S_\delta (q_1,q_2) \nonumber \\
&& \hskip  1cm  \times \sum_{a=1}^3 \tilde\varpi_a  (q_1, q_2)
\biggl [ \langle T(\tilde p_a)\rangle + \tilde B_2(\tilde p_a) +
\tilde B_{3/2}(\tilde p_a) \biggr ]\, , \label{eq65}  \\
{\cal X}_4 &=& {\zeta ^1 \zeta ^2 \over 8 \pi ^2} S_\delta
(q_1,q_2) \sum _{a=1}^3 \biggl [ \partial_{p_a} \partial_{q_1} \ln
E(p_a,q_1) \varpi^*_a(q_2) \nonumber \\
& & \hskip  1cm+ \partial_{p_a}
\partial_{q_2} \ln E(p_a,q_2) \varpi ^*_a(q_1) \biggr ]\, ,
 \\
{\cal X}_5 &=& {\zeta ^1 \zeta ^2 \over 16 \pi ^2} \sum_{a=1}^3
\biggl
[ S_\delta (p_a, q_1) \partial_{p_a} S_\delta (p_a,q_2) \nonumber \\
& & \hskip  1cm - S_\delta (p_a, q_2) \partial_{p_a} S_\delta
(p_a,q_1) \biggr ] \varpi_a (q_1,q_2) \, .
\end{eqnarray}
Furthermore, $\tilde B_2$ and $\tilde B_{3/2}$ are given by
\begin{eqnarray}
\tilde B_2(w) & = & -2 \sum _{a=1}^3 \partial_{p_a} \partial_w \ln
E(p_a,w) \varpi^*_a (w) \, , \\
\tilde B_{3/2}(w) &=& \sum_{\alpha=1}^2 \biggr(  G_2 (w,q_\alpha)
\partial_{q_\alpha} \psi^*_\alpha (q_\alpha) + {3 \over 2}
\partial_{q_\alpha} G_2 (w,q_\alpha) \psi^*_\alpha (q_\alpha)
\biggr)  \, .
\end{eqnarray}
In comparing with \cite{DHokerPhong4} we have written ${\cal
X}_2$, ${\cal X}_3$ together and we didn't split $T(w)$ into
different contributions. We also note that in eq.~(\ref{eq65}) the
three arbitrary points $\tilde p_a$ ($a=1,2,3$) can be different
from the three insertion points $p_a$'s of the $b$ ghost field.
The symbol $\tilde\varpi_a$ is obtained from $\varpi_a$ by
changing $p_a$'s to $\tilde p_a$'s. In the following computation
we will take the limit of $\tilde p_1 \to q_1$. In this limit we
have $\tilde\varpi_{2,3}(q_1,q_2) = 0$ and $\tilde \varpi_1 (q_1,
q_2) = -1$ as we noted at the end of last section. This choice
greatly simplifies the formulas and also make the summation over
spin structure doable (see below and \cite{AllZhu2}).

\section{The chiral measure in hyperelliptic language}

In this section we will compute explicitly the chiral measure in
hyperelliptic language. Here we will not only compute the spin
structure dependent parts, but also compute explicitly all the
quantities appearing in the chiral measure. In order to do this we
will write the chiral measure in hyperelliptic language and then
take the limit of $\tilde p_1 \to q_1$.

Let's first start with ${\cal X}_5$. We have
\begin{equation}
 S(z,q_1)\partial_zS(z,q_2) -
S(z,q_2)\partial_zS(z,q_1)  = {i \over 4 (z-q)^2} S_1(z) .
\end{equation}
The exact phase of the above expression is determined by taking
the limit $z \to q_1$. So the chiral integrand of the
four-particle amplitude (before the summation over spin
structures) from ${\cal X}_5$ is\footnote{An overall factor
${\zeta^1\zeta^2 \over 16 \pi^2 }$ was omitted.}
\begin{equation}
{\cal A}_5 = {i \over 4} \sum_{a=1}^3  { \varpi_a (q_1,q_2) \over
(q-p_a)^2} S_1(p_a) \langle \prod_{i=1}^4 V_i(k_i, \epsilon_i;
z_i,\bar z_i) \rangle \,  ,
\end{equation}
after including the vertex operators.

For ${\cal X}_4$, we need to combine it with the contribution of
$\tilde B_2(q_1)$ in ${\cal X}_2 + {\cal X}_3$ in order to get a
simpler expression. We have
\begin{eqnarray}
  {\cal A}_4 + \tilde B_2(q_1) ~\hbox{in}~ {\cal X}_2 + {\cal
X}_3 & =  & 2 \sum_{a=1}^3  \left[ \partial_{p_a}\partial_{q_1}
\ln E(p_a,q_1) + \partial_{p_a} \partial_{q_2}  \ln E(p_a,q_2)
\right]  \nonumber \\
 & & \times \varpi ^*_a(q_1)   \, S_1(q_1,q_2) \langle \prod_{i=1}^4
V_i(k_i, \epsilon_i; z_i,\bar z_i) \rangle \nonumber \\
 &    & \hskip -3cm = \sum_{a=1}^3 { 2 \over (q-p_a)^2} \, \varpi
^*_a(q_1) \, S_1(q_1,q_2) \langle \prod_{i=1}^4 V_i(k_i,
\epsilon_i; z_i,\bar z_i) \rangle \, , \label{eq39}
\end{eqnarray}
where we have used eq.~(\ref{eq29}) in Appendix A and the fact
$\varpi^*(q_2) = - \varpi^*(q_1)$.

For the rest terms in ${\cal X}_2 + {\cal X}_3$, we first compute
the various contributions from the different fields in the stress
energy tensor. The total stress energy tensor is:
\begin{eqnarray}
T(z) & = & - {1\over 2}: \partial_zX(z)\cdot \partial_zX(z): +
{1\over 2} :\psi(z)\cdot\partial_z\psi(z): \nonumber \\
& &  - :( \partial b c + 2 b\partial c + {1\over
2}\partial\beta\gamma + {3\over 2}\beta\partial\gamma)(z):
\nonumber \\
& \equiv & T_X(z) + T_{\psi}(z) + T_{bc}(z) + T_{\beta\gamma}(z)
\, ,
\end{eqnarray}
in an obvious notations. The various contributions are (following
the notations of \cite{DHokerPhong3}):
\begin{eqnarray}
T_X(w) & = & -10 T_1(w), \label{eq45} \\
T_{\psi}(w) & = & 5 \tilde g_{1/2}(w) =  {5 \over 32}\, (S_1(w))^2,
\\
T_{bc}(w) & = & \tilde g_2(w) - 2 \partial_wf_2(w), \\
T_{\beta\gamma}(w) & = & -\tilde g_{3/2}(w) + {3\over 2}\partial_w
f_{3/2}(w),
\end{eqnarray}
where
\begin{eqnarray}
f_2(w) & = & -{3\over 4} \, \Delta_1(w) +
\sum_{a=1}^3 H(w,p_a) \varpi_a(w,w),  \\
\tilde g_2(w) & = & {5 \over 16} \Delta^2_1(w)  + {3\over 8}\,
\Delta_2(w) \nonumber \\
&  & \hskip -1.5cm + \sum_{a=1}^3 H(w,p_a) \varpi_a(w,w) \left(
{ 1\over w-p_{a+1}}+{1\over w-p_{a+2}} - \Delta_1(w) \right) ,\\
f_{3/2}(w) & = & {\Omega'(w)\over \Omega(w)} + {\Omega(q_1)\over
\Omega(w)} \, S(w,q_1)\psi^*_1(w) + {\Omega(q_2)\over \Omega(w)}
\, S(w,q_2)\psi^*_2(w),\\
\tilde g_{3/2}(w) & = & {1\over 2}\,{\Omega''(w)\over \Omega(w)}
+{1\over 32}\, (S_1(w))^2 \nonumber \\
& & +{\Omega(q_1)\over \Omega(w)} \, S(w,q_1)\partial\psi^*_1(w) +
{\Omega(q_2)\over \Omega(w)} \, S(w,q_2)\partial\psi^*_2(w).
\end{eqnarray}

As we said in the last section we will take the limit of $w\to
q_1$. In this limit $T_{\beta\gamma}(w)$ is singular and we have
the following expansion:
\begin{equation}
T_{\beta\gamma}(w)    =   - { 3/2\over (w-q_1)^2} -
{\partial\psi^*_1(q_1)\over w-q_1} -{1\over8}\Delta_1^2(q)  - {
1\over 32}S_1^2(q) + O(w-q_1).
\end{equation}
The explicit dependence of $G_{3/2}$ on the abelian differential
$\Omega(z)$ drops out in the above expression. These singular
terms are cancelled by similar singular terms in $\tilde
B_{3/2}(w)$. By explicit computation we have:
\begin{eqnarray}
& & \tilde B_{3/2}(w)   =     { 3/2\over (w-q_1)^2} +
{\partial\psi^*_1(q_1)\over w-q_1}   - {1\over 4}\Delta_1^2(q) + {
3\over 4}\Delta_2(q) \nonumber \\
& & \qquad  - \left( {1\over p_1-q} \, { (q-p_2)(q-p_3) \over
(p_1-p_2)(p_1-p_3)} \, \Delta_1(q) + ... \right)
\nonumber \\
& & \qquad - { 3\over 2} \left( {1\over (p_1-q)^2} \,
{(q-p_2)(q-p_3) \over (p_1-p_2)(p_1-p_3)} + ... \right) +
O(w-q_1).
\end{eqnarray}
where $...$ indicates two other terms obtained by cyclicly
permutating $(p_1,p_2,p_3)$. By using the above explicit result we
see that the combined contributions of $T_{\beta\gamma}(w)$ and
$\tilde B_{3/2}(w)$ are non-singular in the limit of $w\to q_1$.
We can then take $\tilde p_1\to q_1$ in ${\cal X}_2 + {\cal X}_3$.
In this limit only $a=1$ contributes to ${\cal X}_2+{\cal X}_3$.
This is because $\tilde\varpi_{2,3}(q_1,q_2) = 0$ and $\tilde
\varpi_1(q_1,q_2) = -1$.  $T_{bc}(w)$ is regular in this limit and
it is spin structure independent. It is given as follows:
\begin{eqnarray}
& & T_{bc}(q_1) = {5\over 16} \Delta_1^2(q_1) - {9\over 8}
\Delta_2(q_1) + \sum_{a=1}^3 { \varpi_a^*(q_1) \over (q_1-p_a)^2}
\nonumber \\
& & \quad + \left\{ {1\over (q_1-p_1)^2}\left[
 1  - {1\over 2} (q_1-p_1) \left( { 1\over q_1-p_2} +
  { 1\over q_1-p_2}
 + \Delta_1(q_1) \right) \right] \right.
 \nonumber \\
 & & \qquad \times \left.  {(q_1-p_2)(q_1-p_3)
 \over(p_1-p_2)(p_1-p_3) } +
 ... \right\} .
 \end{eqnarray}
where ``$...$" again indicates two other terms obtained by
cyclicly permutating $(p_1,p_2,p_3)$.  Combining all the above
results together, we have
\begin{eqnarray}
{\cal A}_2 + {\cal A}_3 & = & -2 S(q_1,q_2) \left\{
 \langle ( T_X(q_1) + T_\psi(q_1) )  \prod_{i=1}^4 V_i
 \rangle_c \right. \nonumber \\
& & + \left[ {1 \over 8}\, (S_1(q))^2 - { 1\over 16} \,
\Delta_1^2(q) - { 3\over 8} \, \Delta_2(q) + \sum_{a=1}^3
{\varpi_a^*(q_1) \over (q-p_a)^2 } \right.  \nonumber \\
& &   - {1\over 2} \, \left\{ {1\over (q -p_1)^2}\left[
 1  +  (q-p_1) \left( { 1\over q -p_2} +  { 1\over q -p_2}
-\Delta_1(q_1) \right)  \right] \right. \nonumber \\
& & \qquad \times  \left. \left. \left. {(q -p_2)(q
-p_3)\over(p_1-p_2)(p_1-p_3) } +  ... \right\} \, \right] \,
\langle \prod_{i=1}^4 V_i \rangle  \, \right\} ,
\end{eqnarray}
by omitting the contribution of $\tilde B_2(q_1)$ which has been
included in eq.~(\ref{eq39}). The subscript ``c" indicates that
the disconnected contraction of $T_\psi$ with $V_i$ should be
omitted. Also by comparing eq.~(\ref{eq39}) with the above
expression we see that it exactly cancels the term with
$\varpi^*_a(q_1)$. Writing these expression together we have:
\begin{eqnarray}
\sum_{i=2}^4 {\cal A}_i & = & -2 S(q_1,q_2) \left\{
 \langle ( T_X(q_1) + T_\psi(q_1) )  \prod_{i=1}^4 V_i
 \rangle \right. \nonumber \\
& & + \left[ {1 \over 8}\, (S_1(q))^2 - { 1\over 16} \,
\Delta_1^2(q) - { 3\over 8} \, \Delta_2(q)  \right.  \nonumber \\
& &   - {1\over 2} \, \left\{ {1\over (q -p_1)^2}\left[
 1  +  (q-p_1) \left( { 1\over q -p_2} +  { 1\over q -p_2}
-\Delta_1(q_1) \right)  \right] \right. \nonumber \\
& & \qquad \times  \left. \left. \left. {(q -p_2)(q
-p_3)\over(p_1-p_2)(p_1-p_3) } +  ... \right\} \, \right] \,
\langle \prod_{i=1}^4 V_i \rangle  \, \right\} .
\end{eqnarray}

Finally we compute  ${\cal X}_1 + {\cal X}_6$ explicitly. By using
the explicit results given in eqs.~(\ref{eq54})--(\ref{eq55}), we
have:
\begin{eqnarray} {\cal A}_1 + {\cal A}_6 & = & - \langle
\psi(q_1)\cdot\partial X(q_1) \psi(q_2)\cdot\partial X(q_2)
\prod_{i=1}^4 V_i \rangle
\nonumber \\
& & - (\partial_{q_1}G_2(q_1,q_2) +
\partial_{q_2}G_2(q_2,q_1) ) S(q_1,q_2) \langle  \prod_{i=1}^4 V_i
\rangle \nonumber \\
& &   + 2 ( G_2(q_1,q_2) + G_2(q_2,q_1) ) \nonumber \\
& & \times  (\partial\psi_1^*(q_1) S(q_1,q_2) -
\partial_{q_2}S(q_1,q_2) ) \langle \prod_{i=1}^4 V_i \rangle .
\label{eqcal16}
\end{eqnarray}
Here it is important that the factor $\partial\psi^*_1(q_2)$
cancels the factor $S(q_1,q_2)$ appearing in the denominator of
$f^{(1)}_{3/2}(q_2)$. The $G_2$'s appearing in the above
expression can be computed explicitly by using eq.~(\ref{eq6}).
The expressions that we will need are given as follows:
\begin{eqnarray}
& & G_2(q_1,q_2) + G_2(q_2,q_1)  =  - {1\over 2} \, \Delta_1(q)
\nonumber \\
& & \qquad + \left[ { 1\over q-p_1} \, { (q-p_2)(q-p_3) \over
(p_1-p_2)
(p_1-p_3)} + ... \right] , \\
& & \partial_{q_1} G_2(q_1,q_2) + \partial_{q_2} G_2(q_2,q_1)  =
{3\over 8}\, \Delta_1^2(q) + { 1\over 4}\, \Delta_2(q) \nonumber
\\
& & \qquad + \left[ {1\over q-p_1}\, \left( {1\over q-p_2} +
{1\over q-p_3} - \Delta_1(q) \right)  { (q-p_2)(q-p_3) \over
(p_1-p_2) (p_1-p_3)} + ... \right] . \nonumber \\
\end{eqnarray}
Now we turn to the computation of the 4-particle amplitude. The
computation was split into three sections.

\section{The 4-particle amplitude I: the vanishing of the connected
term $\langle \psi(q_1)\psi(q_2) \prod_{i=1}^4 k_i \cdot \psi(z_i)
\, \epsilon_i\cdot \psi(z_i)\rangle_c $ }

For graviton and the antisymmetric tensor we use the following
vertex operator (left part only):
\begin{equation}
V_i(k_i,\epsilon_i, z_i) = ( \epsilon_i\cdot \partial X(z_i) + i
k_i\cdot \psi(z_i) \, \epsilon_i \cdot\psi(z_i) )\,  {\rm e}^{ i
k_i \cdot X(z_i, \bar z_i)} .
\end{equation}
Because the vertex operator doesn't contain any ghost fields, all
terms involving ghost fields can be explicit computed which we
have done in the above. For the computation of amplitudes of other
kinds of particles (like fermions), one either resorts to
supersymmetry or can use similar method which was used in
\cite{Sen,Zhu2} to compute the fermionic amplitude.

Before we do explicitly computations, we will show that the
connected term $\langle \psi(q_1)\psi(q_2) \prod_{i=1}^4 k_i \cdot
\psi(z_i) \, \epsilon_i\cdot \psi(z_i)\rangle_c $ gives vanishing
contributions by doing summation over spin structures. This result
was first discovered in \cite{IengoZhu2, Zhu}. Let's recall the
argument here.

First we note that all the connected contractions are the
following three kinds:
\begin{eqnarray}
A_\delta & = & \langle\psi(q_1)\psi(z_1)\rangle
\langle\psi(z_1)\psi(q_2)\rangle \nonumber \\
& & \times \langle\psi(z_2)\psi(z_3)\rangle
\langle\psi(z_3)\psi(z_4)\rangle \langle\psi(z_4)\psi(z_2)\rangle
\,  , \\
B_\delta & = & \langle\psi(q_1)\psi(z_1)\rangle
\langle\psi(z_1)\psi(z_2)\rangle \langle \psi(z_2)\psi(q_2)\rangle
(\langle\psi(z_3)\psi(z_4)\rangle)^2 \, ,
\\
C_\delta & = & \langle\psi(q_1)\psi(z_1)\rangle
\langle\psi(z_1)\psi(z_2)\rangle \nonumber \\
& & \times \langle \psi(z_2) \psi(z_3)\rangle
\langle\psi(z_3)\psi(z_4)\rangle \langle\psi(z_4) \psi(q_2)\rangle
\, ,\\
\end{eqnarray}
or sometimes the expressions permutated among $z_1,z_2,z_3$ and
$z_4$. By using the explicit expression of $\langle \psi(z)
\psi(w) \rangle$ and noting $u(q_2) = - u(q_1)$, we have
\begin{eqnarray}
A_\delta & \propto & 2 \left[ {u(q_1) \over u(z_1)} - {u(z_1)
\over u(q_1)} \right] + \sum_{i\neq j=2}^4 \left[  {u(q_1) u(z_i)
\over u(z_1) u(z_j) } - {u(z_1)u(z_j) \over u(q_1) u(z_i)} \right]
\, ,
\\
B_\delta  & \propto & 2 \sum_{i=1}^2 \left[ {u(q_1) \over u(z_i)}
- {u(z_i) \over u(q_1)} \right] + 2 \left[ {u(z_1) \over u(z_2)} -
{u(z_2) \over u(z_1)} \right] \nonumber \\
& & + \sum_{i=1}^2 \sum_{k\neq l=3}^4 \left[ {u(q_1) u(z_k) \over
u(z_i) u(z_l)} - {u(z_i) u(z_l) \over u(q_1) u(z_k) } \right]
\nonumber \\
& & + \sum_{k\neq l=3}^4 \left[ {u(z_1) u(z_k) \over u(z_2)
u(z_l)} - {u(z_2) u(z_l) \over u(z_1) u(z_k) } \right] \, ,
\\
C_\delta  & \propto &  \sum_{i=1}^4 \left[ {u(q_1) \over u(z_i)} -
{u(z_i) \over u(q_1)} \right] +   \sum_{i < j=1}^4 \left[ {u(z_i)
\over u(z_j)} - {u(z_j) \over u(z_i)} \right] \nonumber \\
& & +  \left[ {u(z_1) u(z_3) \over u(z_2) u(z_4)} - {u(z_2) u(z_4)
\over u(z_1) u(z_3) } \right] + \sum_{i=3}^4   \left[ {u(q_1)
u(z_2) \over u(z_1) u(z_i)} - {u(z_1) u(z_i) \over u(q_1) u(z_2) }
\right]  \nonumber \\
& & + \sum_{i=1}^2 \left[ {u(q_1) u(z_3) \over u(z_i) u(z_4)} -
{u(z_i) u(z_4) \over u(q_1) u(z_3) } \right] \, .
\end{eqnarray}
By using the following ``vanishing identities":
\begin{eqnarray}
& & \sum_{\delta} \eta_\delta Q_\delta \left[ {u(z ) \over u(w)} -
{u(w) \over u(z )} \right] = 0, \\
& & \sum_{\delta} \eta_\delta Q_\delta  \left[ {u(z_1) u(z_2)
\over u(z_3) u(z_4)} - {u(z_3) u(z_4) \over u(z_1) u(z_2) }
\right]  = 0 ,
\end{eqnarray}
given in \cite{AllZhu2}, we readily show that
\begin{equation}
  \sum_{\delta} \eta_\delta Q_\delta   A_\delta =
\sum_{\delta} \eta_\delta Q_\delta   B_\delta =\sum_{\delta}
\eta_\delta Q_\delta   C_\delta = 0.
\end{equation}
This shows that all the connected terms in $\langle
\psi(q_1)\psi(q_2) \prod_{i=1}^4 k_i \cdot \psi(z_i) \,
\epsilon_i\cdot \psi(z_i)\rangle_c $ give vanishing contributions
to the 4-particle amplitude  after summation over spin structures,
and can be neglected. For the first term in eq.~(\ref{eqcal16}) we
only need to compute the disconnected term $S(q_1,q_2)\langle
\partial X(q_1) \cdot \partial X(q_2) \prod_{i=1}^4 V_i\rangle$.

\section{The 4-particle amplitude II: the $S_n(q) \langle
\prod_{i=1}^4 k_i \cdot \psi(z_i) \, \epsilon_i\cdot
\psi(z_i)\rangle $ terms}

The next step in the explicit calculation of the 4-particle
amplitude is to compute  the terms with a single $S_n(q)$ (or
$S_1(p_a)$ from ${\cal A}_5$) in the summation over spin
structures, i.e. terms like $S_n(q) \langle \prod_{i=1}^4 k_i
\cdot \psi(z_i) \, \epsilon_i\cdot \psi(z_i)\rangle$. Other terms
with less $\psi$ fields automatically give  0 after summation over
spin structures (the non-renormalization theorem verified in
\cite{AllZhu2}).

The first step in this computation is to do the contraction for
the $\psi$ fields. We have
\begin{eqnarray}
\langle \prod_{i=1}^4 k_i\cdot\psi(z_i) \epsilon_i\cdot\psi(z_i)
\rangle & = &   K_1(1,2,3,4)(S(z_1,z_2)S(z_3,z_4))^2 \nonumber \\
& & + K_1(1,3,4,2)(S(z_1,z_3)S(z_2,z_4))^2 \nonumber \\
& & + K_1(1,4,2,3)(S(z_1,z_4)S(z_2,z_3))^2 \nonumber \\
& & + K_2(1,2,3,4) S(z_1,z_2 ,z_3 ,z_4) \nonumber \\
& & + K_2(1,3,4,2) S(z_1,z_3 ,z_4 ,z_2)  \nonumber \\
& & + K_2(1,4,2,3) S(z_1,z_4 ,z_2 ,z_3 )   \, , \label{eq68}
\end{eqnarray}
where
\begin{eqnarray}
S(z_1,z_2,z_3,z_4) & = & S(z_1,z_2)S(z_2,z_3) S(z_3,z_4)
S(z_4,z_1) , \\
 K_1(1,2,3,4) & = & (k_1\cdot k_2 \,
\epsilon_1\cdot\epsilon_2
-k_1\cdot \epsilon_2\, k_2\cdot\epsilon_1) \nonumber \\
& &   \times (k_3\cdot k_4 \, \epsilon_3\cdot\epsilon_4
-k_3\cdot \epsilon_4\, k_4\cdot\epsilon_3), \\
K_2(1,2,3,4) & = & -k_1\cdot k_2 k_3\cdot k_4
\epsilon_1\cdot\epsilon_4
 \epsilon_2\cdot\epsilon_3 - k_1\cdot k_4 k_2\cdot k_3
 \epsilon_1\cdot\epsilon_2  \epsilon_3\cdot\epsilon_4 \nonumber \\
  & &
 \hskip -2.5cm+ k_1\cdot k_2(k_3\cdot\epsilon_4
 k_4\cdot\epsilon_1\epsilon_2\cdot\epsilon_3 +k_3\cdot\epsilon_2
 k_4\cdot\epsilon_3\epsilon_1\cdot\epsilon_4 - k_3\cdot\epsilon_2
 k_4\cdot\epsilon_1\epsilon_3\cdot\epsilon_4) \nonumber \\
 & &
 \hskip -2.5cm+ k_1\cdot k_4(k_2\cdot\epsilon_1
 k_3\cdot\epsilon_2\epsilon_3\cdot\epsilon_4 +k_2\cdot\epsilon_3
 k_3\cdot\epsilon_4\epsilon_1\cdot\epsilon_2 - k_2\cdot\epsilon_1
 k_3\cdot\epsilon_4\epsilon_2\cdot\epsilon_3) \nonumber \\
  & &
 \hskip -2.5cm+ k_2\cdot k_3(k_1\cdot\epsilon_2
 k_4\cdot\epsilon_1\epsilon_3\cdot\epsilon_4 +k_1\cdot\epsilon_4
 k_4\cdot\epsilon_3\epsilon_1\cdot\epsilon_2 - k_1\cdot\epsilon_2
 k_4\cdot\epsilon_3\epsilon_1\cdot\epsilon_4)\nonumber \\
  & &
 \hskip -2.5cm+ k_3\cdot k_4(k_1\cdot\epsilon_2
 k_2\cdot\epsilon_3\epsilon_1\cdot\epsilon_4 +k_1\cdot\epsilon_4
 k_2\cdot\epsilon_1\epsilon_2\cdot\epsilon_3 - k_1\cdot\epsilon_4
 k_2\cdot\epsilon_3\epsilon_1\cdot\epsilon_2) \nonumber \\
  & &
 \hskip -2.5cm- k_1\cdot\epsilon_2 k_2\cdot\epsilon_3 k_3
 \cdot\epsilon_4
k_4\cdot\epsilon_1 -k_1\cdot\epsilon_4 k_2\cdot\epsilon_1
k_3\cdot\epsilon_2 k_4\cdot\epsilon_3 .
\end{eqnarray}
We note  that the above two kinematic factors have a symmetry
under the simultaneous interchange of $1 \leftrightarrow 4$ and $2
\leftrightarrow 3$. $K_2$ is also invariant under the cyclic
permutations of $(1,2,3,4)$, i.e., $ K_2(1,2,3,4) = K_2(2,3,4,1) =
K_2(3,4,1,2)$, etc.

By using the explicit expression for the Szeg\"o kernel
$S(z_1,z_2)$ given in eq.~(\ref{eqszego}), we have
\begin{eqnarray}
(S(z_1,z_2)S(z_3,z_4))^2  & = & { 1\over 16\, z_{12}^2 z_{34}^2
}\left\{ 4 + 2\left[ {u(z_1) \over u(z_2)} + {u(z_2) \over u(z_1)}
\right] + 2\left[ {u(z_3) \over u(z_4)} +
{u(z_4) \over u(z_3)} \right]  \right. \nonumber \\
& & \hskip -2.5cm \left.  + \left[ {u(z_1) u(z_3) \over u(z_2)
u(z_4) } + {u(z_2) u(z_4) \over u(z_1) u(z_3) } \right]  + \left[
{ u(z_1) u(z_4) \over u(z_2) u(z_3) } + {u(z_2) u(z_3) \over
u(z_1) u(z_4)} \right] \right\} ,
\\
S(z_1,z_2,z_3,z_4) & = & {1\over 16\, z_{12}z_{23}z_{34}z_{41} }
\left\{ 2 + \sum_{i<j=1}^4 \left[ {u(z_i) \over u(z_j)} + {u(z_j)
\over u(z_i)} \right] \right.  \nonumber \\
& &  \qquad + \left.    \left[ {u(z_1) u(z_3) \over u(z_2) u(z_4)
} + {u(z_2) u(z_4) \over u(z_1) u(z_3) } \right]  \right\}.
\end{eqnarray}
By using the ``vanishing identities" proved in \cite{AllZhu2},
most of the terms are 0 after summation over spin structures. In
order to do the summation over spin structure  for the rest
non-vanishing expression we need the following summation formula:
\begin{eqnarray}
& &  \sum_\delta \eta_\delta Q_\delta \left\{ {u(z_1)  u(z_2)
\over u(z_3)u(z_4)} - (-1)^n {u(z_1)u(z_2)\over u(z_3)u(z_4) }
\right\} (S_m(x))^n \nonumber  \\
& &  \qquad \qquad =  {2 P(a) \prod_{i=1}^2\prod_{j=3}^4 (z_i-z_j)
\prod_{i=1}^4(x-z_i) \over y^2(x) \prod_{i=1}^4 y(z_i) } \times
C_{n,m}, \label{eq999}
\end{eqnarray}
where
\begin{eqnarray}
C_{1,1} & = & 1,  \label{eq991} \\
C_{2,1} & = & - 2 (\tilde z_1 + \tilde z_2 - \tilde z_3 - \tilde
z_4) ,     \\
C_{1,2} & = & \Delta_1(x)  - \sum_{k=1}^4 \tilde z_k   , \\
C_{3,1} & = & 2 \Delta_2(x) - \Delta_1^2(x) + 2 \Delta_1(x)\,
\sum_{k=1}^4 \tilde z_k \nonumber \\
& &  + 4 \left(\tilde z_1 \tilde z_2   - 2 ( \tilde z_1  + \tilde
z_2 )(\tilde z_3 + \tilde z_4 ) +
\tilde z_3  \tilde z_4  \right), \label{eq992} \\
\tilde z_k & = & { 1\over x - z_k}, \\
 P(a) & = & \prod_{i<j}(a_i-a_j).
\end{eqnarray}
$C_{1,1}$ and $C_{1,2}$ were firstly derived in \cite{IengoZhu2}
(see also \cite{Zhu}). Although other values of $n,m$ also gives
modular invariant expressions, the results are quite
complex.\footnote{This is due to the non-vanishing of the
summation over spin structures when we set $z_1=z_3$ or $z_1=z_4$,
etc.} Fortunately we only need to use the above listed results.
The proof of the above summation formula can be found in Appendix
B.

By using these result one easily prove the following:
\begin{eqnarray}
& & \hskip -2cm \sum_{\delta}\eta_\delta Q_\delta  \, S_n(x)
S(z_1,z_2,z_3,z_4) = \sum_{\delta}\eta_\delta Q_\delta  \, S_n(x)
(S(z_1,z_2)S(z_3,z_4))^2 \nonumber \\
& =  & { P(a) \over 8 \, y^2(x) }\, \prod_{i=1}^4 { x-z_i \over
y(z_i)} \, \left\{
\begin{array}{ll}
1,  & n =1, \\
\Delta_1(x) - \sum_{i=1}^4 \tilde z_i, & n=2.
\end{array}
\right.
\end{eqnarray}
By using this result and eq.~(\ref{eq68}), we have
\begin{eqnarray}
& & \sum_{\delta}\eta_\delta Q_\delta  \, S_n(x)\, \langle
\prod_{i=1}^4 k_i\cdot\psi(z_i) \epsilon_i\cdot\psi(z_i) \rangle
\\
& = & K(k_i, \epsilon_i) \, { P(a) \over 8\,  y^2(x) }\,
\prod_{i=1}^4 { x-z_i \over y(z_i)} \times \left\{
\begin{array}{ll}
1,  & n =1, \\
\Delta_1(x) - \sum_{i=1}^4 \tilde z_i, & n=2,
\end{array} \right.
\end{eqnarray}
where
\begin{eqnarray}
 K(k_i, \epsilon_i)  & \equiv &  K_1(1,2,3,4) + K_1(1,3,4,2)
+ K_1(1,4,2,3) \nonumber \\
& &  + K_2(1,2,3,4)  + K_2(1,3,4,2)  + K_2(1,4,2,3)   \nonumber \\
& = & -{1\over 4} (s\, t\epsilon_1\cdot\epsilon_3
\epsilon_2\cdot\epsilon_4 +s\, u \epsilon_2\cdot\epsilon_3
\epsilon_1\cdot\epsilon_4 + t\, u \epsilon_1\cdot\epsilon_2
\epsilon_3\cdot\epsilon_4 )
\nonumber \\
& & + {1\over 2} \, s \, ( \epsilon_1\cdot k_4 \epsilon_3\cdot k_2
\epsilon_2 \cdot \epsilon_4  + \epsilon_2\cdot k_3 \epsilon_4\cdot
k_1 \epsilon_1 \cdot \epsilon_3
\nonumber \\
& & \qquad + \epsilon_1\cdot k_3 \epsilon_4\cdot k_2 \epsilon_2
\cdot \epsilon_3  + \epsilon_2\cdot k_4 \epsilon_3\cdot k_1
\epsilon_1 \cdot \epsilon_4  )
\nonumber \\
& & + {1\over 2} \, t \, ( \epsilon_2\cdot k_1 \epsilon_4\cdot k_3
\epsilon_1 \cdot \epsilon_3  + \epsilon_3\cdot k_4 \epsilon_1\cdot
k_2 \epsilon_2 \cdot \epsilon_4
\nonumber \\
& & \qquad + \epsilon_2\cdot k_4 \epsilon_1\cdot k_3 \epsilon_3
\cdot \epsilon_4  + \epsilon_3\cdot k_1 \epsilon_4\cdot k_2
\epsilon_1 \cdot \epsilon_2  )
\nonumber \\
& & + {1\over 2} \, u \, ( \epsilon_1\cdot k_2 \epsilon_4\cdot k_3
\epsilon_2 \cdot \epsilon_3  + \epsilon_3\cdot k_4 \epsilon_2\cdot
k_1 \epsilon_1 \cdot \epsilon_4
\nonumber \\
& & \qquad + \epsilon_1\cdot k_4 \epsilon_2\cdot k_3 \epsilon_1
\cdot \epsilon_4  + \epsilon_3\cdot k_2 \epsilon_4\cdot k_1
\epsilon_1 \cdot \epsilon_2  ),
\end{eqnarray}
is the standard kinematic factor (left part only) appearing at
tree and one loop computation of superstring theories
\cite{GreenSchwarz1, IengoZhu2, Zhu}. So apart from an  overall
factor and the kinematic factor, each appearance of $  S_n(x)\,
\langle \prod_{i=1}^4 V_i(k_i,\epsilon;z_i,\bar z_i) \rangle$  in
the chiral integrand ${\cal A}_i$ can be substituted by either $1$
or $\Delta_1(x) - \sum_{i=1}^4 \tilde z_i$ for $n=1$ or 2. After
this has been done we have the following form for the chiral
integrand:\footnote{omitting the $(S_1(q))^3$ and $T_\psi$ terms
and an overall factor ${ P(a) \over 8 y^2(q)} \, \prod_{i=1}^4 {
q- z_i \over y(z_i)}$. So the contribution from ${\cal A}_5$ has
an extra factor of ${y^2(q)\over y^2(p_a)}\, \prod_{i=1}^4
{p_a-z_i \over q-z_i}$. $\tilde z_k$ in the following denotes ${
1\over q - z_k}$. }
\begin{eqnarray}
{\cal A}  & = & \sum_{a=1}^3 {\varpi_a(q_1,q_2) \over (q-p_a)^2 }
\, \times \, {y^2(q)\over y^(p_a)}\, \prod_{i=1}^4 {p_a-z_i \over
q-z_i}  \, \langle \prod_{i=1}^4 {\rm e}^{ i k_i \cdot X(z_i, \bar
z_i)} \rangle \nonumber \\
& &  +  \langle( \partial X(q_1) \cdot ( \partial X(q_1) +
\partial X(q_2) ) \prod_{i=1}^4 \hbox{e}^{i k_i \cdot X(z_i, \bar
z_i)} \rangle \nonumber \\
& & + \left\{ {1\over 8}\Delta_1^2(q) + {3\over 4}\Delta_2(q) +
\left\{  \left[ {1\over (q-p_1)^2}+ {1\over q-p_1} \right. \right.
\right. \nonumber \\
& & \qquad \times \left.  \left. \left( {1\over q-p_2} + {1\over
q-p_3} - \Delta_1(q)\right) \right] \, {(q-p_2)(q-p_3)\over
(p_1-p_2)(p_1-p_3)} +\cdots \right\}
\nonumber \\
& & - { 3\over 8}\Delta^2_1(q) - {1\over 4}\Delta_2(q) - \left[
{1\over q-p_1}\left(  {1\over q-p_2} + {1\over q-p_3} -
\Delta_1(q)\right)  \right. \nonumber \\
& & \quad \left. \times  {(q-p_2)(q-p_3)\over (p_1-p_2)(p_1-p_3)}
+\cdots \right]  + { 1\over 2}\Delta_1(q) \sum_{k=1}^4\tilde z_k
\nonumber \\
& & \left.- \left[ {1\over q-p_1}  \, {(q-p_2)(q-p_3)\over
(p_1-p_2)(p_1-p_3)} +\cdots \right] \sum_{k=1}^4 \tilde z_k
\right\} \,  \langle \prod_{i=1}^4 {\rm e}^{ i k_i \cdot X(z_i,
\bar z_i)} \rangle \nonumber \\
& = & \langle :\partial X(q_1) \cdot ( \partial X(q_1) +
\partial X(q_2) ): \prod_{i=1}^4 \hbox{e}^{i k_i \cdot X(z_i, \bar
z_i)} \rangle \nonumber \\
& & +\left[ {1\over 2}\Delta_2(q) - { 1\over 4}\Delta_1^2(q) + {
1\over 2}\Delta_1(q)\sum_{i=1}^4 \tilde z_k - \sum_{k<l} \tilde
z_k \tilde z_l\right] \, \langle\prod_{i=1}^4 \hbox{e}^{i k_i
\cdot X(z_i, \bar z_i)} \rangle, \nonumber \\
\end{eqnarray}
and the complete chiral integrand is
\begin{eqnarray}
{\cal A}  & = & K(k_i,\epsilon_i) \left\{  \langle : \partial
X(q_1) \cdot ( \partial X(q_1) + \partial X(q_2) ) : \prod_{i=1}^4
\hbox{e}^{i k_i \cdot X(z_i, \bar z_i)} \rangle  \right. \nonumber \\
& & + \left[ {1\over 2}\Delta_2(q) - { 1\over 4}\Delta_1^2(q) + {
1\over 2}\Delta_1(q)\sum_{i=1}^4 \tilde z_k - \sum_{k<l} \tilde
z_k \tilde z_l\right] \nonumber \\
& & \quad \times \left.  \langle\prod_{i=1}^4 \hbox{e}^{i k_i
\cdot X(z_i, \bar z_i)} \rangle \right\} \times {P(a)\over 8\,
y^2(q)} \, \prod_{i=1}^4 {q-z_i \over y(z_i)}\nonumber  \\
& &  - \left\langle\left\langle {1\over 4}(S_1(q))^3  \langle
\prod_{i=1}^4 V_i \rangle  +   S_1(q) \,  \langle :\psi(q_1)\cdot
\partial \psi(q_1): \prod_{i=1}^4 V_i \rangle_c
\right\rangle\right\rangle_s. \label{eq00}
\end{eqnarray}
Here we have reinserted the overall factor. The subscript $c$
indicates that the self contraction of $\psi(q_1)$ with
$\partial\psi(q_1)$ is omitted and $\langle \langle \cdots \rangle
\rangle_s$ indicates that appropriate summation over spin
structure should be carried out. As one can see from the above
results, all the dependence on $p_a$'s drops out. This is a very
strong check for the validity of the new supersymmetric gauge
fixing method at two loops.

\section{The 4-particle amplitude III: terms from $(S_1(q))^3$
and $T_\psi$}

To compute the last term in eq.~(\ref{eq00}), we need the
following summation formulas:
\begin{eqnarray}
& & \sum_\delta \eta_\delta Q_\delta (
S(x,z_1)S(z_1,z_2)\partial_xS(z_2,x) + (z_1\leftrightarrow z_2))
\, (S(z_3,z_4))^2 \, S_1(x)   \nonumber \\
& & \qquad = - {P(a)\over 16\,  y^2(x) }\, \prod_{i=1}^4{
x-z_i\over  y(z_i)} \, \left( \tilde z_{14}\tilde z_{23} + \tilde
z_{13}\tilde z_{24} \right) , \label{eq87}
\\
& & \sum_\delta \eta_\delta Q_\delta (
S(x,z_1)S(z_1,z_2)S(z_2,z_3)S(z_3,z_4)\partial_xS(z_4,x) \nonumber
\\
& &  \qquad \quad + (z_1\leftrightarrow z_4, z_2\leftrightarrow
z_3)) \, \, S_1(x)   =  {P(a)\over 16\,  y^2(x)}\, \prod_{i=1}^4{
x-z_i\over  y(z_i)} \,  \tilde z_{14} \tilde z_{23} ,  \label{eq88}\\
& & \sum_\delta \eta_\delta Q_\delta  (S(z_1,z_2)S(z_3,z_4))^2
(  S_1(x)  )^3 \nonumber \\
& & \qquad =  {P(a)  \over 8\, y^2(x)}\, \prod_{i=1}^4{ x-z_i\over
y(z_i)} \, \left[ 2 \Delta_2(x) - \Delta_1^2(x) +
2 \Delta_1(x)\sum_{i=1}^4{\tilde z_i}  \right. \nonumber \\
& & \qquad \qquad \left. + 4  \sum_{k<l}  \tilde z_k\tilde z_l   -
24( { \tilde z_1\tilde z_2 } +{ \tilde z_3\tilde z_4 } ) \right],
\label{eq89}
\\
& & \sum_\delta \eta_\delta Q_\delta
S(z_1,z_2)S(z_2,z_3)S(z_3,z_4)S(z_4,z_1) ( S_1(x)  )^3 \nonumber \\
& & \qquad    =  {P(a) \over 8 \, y^2(x)}\, \prod_{i=1}^4{
x-z_i\over  y(z_i)} \, \left[ 2 \Delta_2(x) - \Delta_1^2(x) +
2 \Delta_1(x)\sum_{i=1}^4{\tilde z_i}  \right. \nonumber \\
& & \qquad \qquad \left. + 4  \sum_{k<l}  \tilde z_k\tilde z_l   -
12( { \tilde z_1+ \tilde z_3 })({ \tilde z_2+\tilde z_4 } )
\right] . \label{eq90}
\end{eqnarray}
All these formulas can be derived by using the formulas given in
eq.~(\ref{eq999}) and the following  useful formula for the
derivative of the Szeg\"o kernel:
\begin{equation}
\partial_x S(z, x) = { 1\over  2(z-x)^2 }
\, {u(z) + u(x) \over \sqrt{u(z)u(x)}  }  - {S_1(x) \over 8 \,
(z-x) } \, {u(z) - u(x) \over \sqrt{u(z)u(x)} } \,   .
\end{equation}

Now we compute the last term explicitly in eq.~({\ref{eq00}).
First we have the following contractions:
\begin{eqnarray}
& & \hskip -1.5cm \langle \psi(x)\cdot\partial\psi(x)
\prod_{i=1}^4
k_i\cdot\psi(z_i) \epsilon_i\cdot\psi(z_i) \rangle \nonumber \\
& & =   -2 \, CS_1(1,2,3,4)K_1(1,2,3,4)   - 2 \,
CS_1(1,3,4,2)K_1(1,3,4,2) \nonumber \\
& & -2 \, CS_1(1,3,4,2)K_1(1,3,4,2)
 +    CS_2(1,2,3,4) K_2(1,2,3,4) \nonumber \\
& &   + CS_2(1,3,4,2)K_2(1,3,4,2)+ CS_2(1,3,4,2)K_2(1,3,4,2) ,
\end{eqnarray}
where
\begin{eqnarray}
CS_1(1,2,3,4) & = & (S(x,z_1)S(z_1,z_2)\partial_xS(z_2,x)+
(z_1\leftrightarrow z_2)) (S(z_3,z_4))^2  \nonumber \\
& &  \hskip -1.5cm + (S(x,z_3)S(z_3,z_4)\partial_xS(z_4,x)+
(z_3\leftrightarrow
z_4)) (S(z_1,z_2))^2 , \\
CS_2(1,2,3,4) & = & S(x,z_1,z_2,z_3,z_4) + S(x,z_2,z_3,z_4,z_1)
\nonumber \\
& & + S(x,z_3,z_4,z_1,z_2) + S(x,z_4,z_1,z_2,z_3) \nonumber \\
& & + S(x,z_3,z_2,z_1,z_4) + S(x,z_4,z_3,z_2,z_1)
\nonumber \\
& & + S(x,z_4,z_3,z_2,z_1)  + S(x,z_4,z_3,z_2,z_1), \\
S(x, z_1,z_2,z_3,z_4) & \equiv &
S(x,z_1)S(z_1,z_2)S(z_2,z_3)S(z_3,z_4)\partial_xS(z_4,x) ,
\end{eqnarray}
and the kinematic factors $K_{1,2}(1,2,3,4)$ are defined as
before. By using eq.~(\ref{eq87}), one sees that $CS_1(1,2,3,4)$
gives a factor:
\begin{eqnarray}
& & - { 1\over 2}( \tilde z_{14} \tilde z_{23} +  \tilde z_{13}
\tilde z_{24} + ( \tilde z_1\leftrightarrow \tilde z_3, \tilde
z_2\leftrightarrow \tilde z_4) ) \nonumber
\\
& & \qquad = - 3 (\tilde z_1\tilde z_2 +\tilde z_3\tilde z_4) +
\sum_{k<l} \tilde z_k \tilde z_l,
\end{eqnarray}
and by using eq.~(\ref{eq88}), one sees that  $CS_1(1,2,3,4)$
gives a factor:
\begin{eqnarray}
& &   { 1\over 2}( \tilde z_{14} \tilde z_{23} + \tilde z_{21}
\tilde z_{34} + \tilde z_{32} \tilde z_{41} + \tilde z_{43} \tilde
z_{12}) \nonumber \\
&  & \qquad =  3 (\tilde z_1+\tilde z_3)(\tilde z_2+\tilde z_4)- 2
\sum_{k<l} \tilde z_k \tilde z_l.
\end{eqnarray}
By using these results one computes the connected contribution of
the $T_\psi$ term as follows:
\begin{eqnarray}
{\cal A}_{T_\psi} & = & -\left[ - 2 (- 3(\tilde z_1\tilde z_2+
\tilde z_3\tilde z_4)+ \sum_{k<l}
\tilde z_k\tilde z_l) K_1(1,2,3,4) \right. \nonumber \\
& & - 2( -3(\tilde z_1\tilde z_3+ \tilde z_2\tilde z_4) +
\sum_{k<l} \tilde z_k\tilde z_l) K_1(1,3,4,2)
\nonumber \\
& & + 2( - 3(\tilde z_1\tilde z_4+ \tilde z_2\tilde z_3) +
\sum_{k<l} \tilde z_k\tilde z_l) K_1(1,4,2,3)
\nonumber \\
& & + ( 3 (\tilde z_1+ \tilde z_3)(\tilde z_2+\tilde z_4) -
2\sum_{k<l} \tilde z_k\tilde z_l) K_2(1,2,3,4)
\nonumber \\
& & + ( 3 (\tilde z_1+ \tilde z_4)(\tilde z_2+\tilde z_3) -
2\sum_{k<l} \tilde z_k\tilde z_l) K_2(1,3,4,2)
\nonumber \\
& & \left. + ( 3 (\tilde z_1+ \tilde z_2)(\tilde z_3+\tilde z_4) -
2 \sum_{k<l} \tilde z_k\tilde z_l) K_2(1,4,2,3) \right] \nonumber
\\
& = & 2 K(k_i,\epsilon_i) \, \sum_{k<l} \tilde z_k\tilde z_l
\nonumber \\
& & - 6  (\tilde z_1\tilde z_2+ \tilde z_3\tilde z_4) K_1(1,2,3,4)
-6 (\tilde z_1\tilde z_3+ \tilde z_2\tilde z_4)   K_1(1,3,4,2)
\nonumber \\
& & -6 (\tilde z_1\tilde z_4+ \tilde z_2\tilde z_3) K_1(1,4,2,3) -
3 (\tilde z_1+ \tilde z_3)(\tilde z_2+\tilde z_4)   K_2(1,2,3,4)
\nonumber \\
& & -3 (\tilde z_1+ \tilde z_4)(\tilde z_2+\tilde z_3)
K_2(1,3,4,2)- 3 (\tilde z_1+ \tilde z_2)(\tilde z_3+\tilde z_4) ,
\label{eqss102}
\end{eqnarray}
apart from an overall factor.

The $ (S_1(q))^3$ contribution can also be computed. We need to
use eq.~(\ref{eq68}) for the $\psi$ contraction and then eqs.
(\ref{eq89}) and (\ref{eq90}) for the summation over spin
structures. The result is
\begin{eqnarray}
{\cal A}_{S_1^3} & = &-{ 1\over 4} \Big[ ( 2 \Delta_2(q) -
\Delta_1^2(q) + 2 \Delta_1(q) \sum_{k} \tilde z_k + 4 \sum_{k<l}
\tilde z_k\tilde z_l
)K(k_i,\epsilon_i)   \nonumber \\
& & - 24 (\tilde z_1 \tilde z_2 + \tilde z_3\tilde z_4)
K_1(1,2,3,4) - 24 (\tilde z_1\tilde z_3 + \tilde z_2
\tilde z_4)K_1(1,3,4,2) \nonumber \\
& & - 24 (\tilde z_1 \tilde z_4 + \tilde z_2\tilde z_3)
K_1(1,4,2,3) \nonumber \\
& & - 12 (\tilde z_1+\tilde z_3)(\tilde z_2 +
\tilde z_4)K_2(1,2,3,4) \nonumber \\
& &   - 12 (\tilde z_1+\tilde z_4)(\tilde z_2 + \tilde
z_3)K_2(1,3,4,2) \nonumber \\
& & - 12 (\tilde z_1+\tilde z_2)(\tilde z_3 + \tilde
z_4)K_2(1,4,2,3) \Big]\, . \label{eqss103}
\end{eqnarray}
By using eq.~(\ref{eqss102}) and eq.~(\ref{eqss103}), we have
\begin{eqnarray}
{\cal A}_{T_\psi} +  {\cal A}_{S_1^3} & = & \left[ - { 1\over 2}
\Delta_2(q) + {1\over 4} \Delta_1^2(q) - {1\over 2}  \Delta_1(q)
\sum_{k} \tilde z_k + \sum_{k<l} \tilde z_k\tilde z_l \right]
\nonumber \\
& & \quad \times K(k_i,\epsilon_i) \, \langle \prod_{i=1}^4
\hbox{e}^{i k_i \cdot X(z_i, \bar z_i)} \rangle  { P(a) \over 8
y^2(q) }\,  \prod_{i=1}^4 { q -z_i\over y(z_i) } \, ,
\label{eq900}
\end{eqnarray}
by including the overall factor. The above contribution also has
the standard kinematic factor and it exactly cancels the second
part of the first term of eq.~(\ref{eq00}).

By using eq.~(\ref{eq00}) and eq.~(\ref{eq900}), we found that
final result for the complete chiral integrand of the
four-particle amplitude is quite simple and it is given as
follows:
\begin{equation}
{\cal A} = K(k_i,\epsilon_i) \langle : \partial X(q_1) \cdot (
\partial X(q_1) +
\partial X(q_2) ) : \prod_{i=1}^4 \hbox{e}^{i k_i \cdot X(z_i, \bar
z_i)} \rangle  \prod_{i=1}^4 { q -z_i\over y(z_i) } , \label{eqxx}
\end{equation}
where  an overall factor $P(a)\over 8 y^2(q)$ is omitted. This
factor is cancelled by an identical factor from the factor ${\cal
Z}$ (see \cite{IengoZhu2} and Appendix C). In the next section we
will combine the above result with the measure and write the
complete expression for the two loop four-particle amplitude. Some
properties of  this chiral integrand will also be studied in the
next section.

\section{The 4-particle amplitude IV: the final result}

The chiral integrand obtained in the last section is not symmetric
under the interchange $q_1 \leftrightarrow q_2$. If we take the
limit of $\tilde p_1 \to q_2$ we would obtain the same result with
$q_1 \leftrightarrow q_2$. So the final result should be a
symmetrization of eq.~(\ref{eqxx}). We suspect that this ambiguity
is caused by our use of the delta function super Beltrami
differentials and the final result actually requires this
symmetrization if we study carefully the limit. We will not do
this in this paper. In fact this symmetrization also makes the
final amplitude explicitly independent of $q$ as we will show
immediately.

The symmetrized chiral integrand is
\begin{eqnarray}
 {\cal A} & = &
K(k_i,\epsilon_i) \langle : (\partial X(q_1) +\partial X(q_2))
\cdot ( \partial X(q_1) +
\partial X(q_2) ) : \nonumber \\
& & \quad \times \prod_{i=1}^4 \hbox{e}^{i k_i \cdot X(z_i, \bar
z_i)} \rangle \prod_{i=1}^4 { q -z_i\over y(z_i) } . \label{eqxx1}
\end{eqnarray}
By using eq.~(\ref{eq27}) for the $\langle \partial X(z) \partial
X(w)\rangle$ correlators we have
\begin{equation}
\langle :  \partial X(q_1)   \cdot ( \partial X(q_1) +
\partial X(q_2) ) : \rangle = 0. \label{eq106}
\end{equation}
To compute explicitly eq.~(\ref{eqxx1}) we also need
eq.~(\ref{eq115}) for the $\langle \partial X(z) X(w,\bar
w)\rangle $ correlators and we have
\begin{equation}
\sum_{i=1}^4 \langle (\partial X^\mu (q_1) +\partial  X^\mu (q_2))
k_i\cdot X(w_i) \rangle = - \sum_{i=1}^4 { k_i^\mu \over q - w_i}
\, . \label{eq107}
\end{equation}
By using  eq.~(\ref{eq106}) and eq.~(\ref{eq107}) in
eq.~(\ref{eqxx1}) we have:
\begin{eqnarray}
{\cal A} & = &   K(k_i,\epsilon_i) \, \left[ i \sum_{i=1}^4 { k_i
\over q - z_i} \right]^2   \prod_{i<j} {\rm exp}\left[ - k_i\cdot
k_j \ G(z_i,z_j) \right]  \,   \prod_{i=1}^4 { q - z_i \over
y(z_i) } \nonumber \\
& = & { K(k_i,\epsilon_i) \over   \prod_{i=1}^4 y(z_i) }  \,
\prod_{i<j} {\rm exp}\left[ - k_i\cdot k_j \ G(z_i,z_j) \right]
 \nonumber \\
&  & \times ( s (z_1z_2 + z_3 z_4) + t(z_1z_4+ z_2 z_3) + u(z_1z_3
+ z_2 z_4)) ,
\end{eqnarray}
which is  independent of  the  insertion points $q_{1,2}$. Here
$G(z_i,z_j)$ is the scalar Green function which is given in terms
of the prime form $E(z_i,z_j)$ as follows (see
\cite{DHokerPhong2}):
\begin{equation}
G(z,w)   =  - \ln |E(z,w)|^2 + 2 \pi {\rm Im}\int_z^w \omega_I
({\rm Im}\Omega)^{-1}_{IJ}  {\rm Im} \int_z^w \omega_J .
\label{eq110}
\end{equation}
$s,t,u$ are the standard Mandelstam variables, $s = -
(k_1+k_2)^2$, etc.

For type II superstring theory the complete integrand is
\begin{eqnarray}
{\cal A}_{II} & = &  K(k_i,\epsilon_i) \langle : (\partial X(q_1)
+\partial X(q_2)) \cdot ( \partial X(q_1) +
\partial X(q_2) ) :  \nonumber \\
& & \times : (\bar{\partial} X(\tilde{\bar q}_1) +{\bar\partial}
X(\tilde{\bar q}_2)) \cdot ( \bar{\partial} X(\tilde{\bar q}_1) +
\bar{\partial} X(\tilde{\bar q}_2) ) :  \nonumber \\
& & \times \prod_{i=1}^4 {\rm e}^{i k_i \cdot X(z_i, \bar z_i)}
\rangle \prod_{i=1}^4 { (q -z_i)(\tilde{\bar q}-\bar z_i)
\over |y(z_i)|^2 } \nonumber \\
&  = &   { K(k_i,\epsilon_i) \over   \prod_{i=1}^4 |y(z_i) |^2 }
\, \prod_{i<j} {\rm exp}\left[ - k_i\cdot k_j \ G(z_i,z_j) \right]
 \nonumber \\
&  & \times | s (z_1z_2 + z_3 z_4) + t(z_1z_4+ z_2 z_3) + u(z_1z_3
+ z_2 z_4)|^2 , \label{eq111}
\end{eqnarray}
which is  independent of the left part insertion points $q_{1,2}$
and also the right part insertion points $\tilde q_{1,2}$.

The amplitude is obtained by integrating over the moduli space. At
two loops, the moduli space can be parametrized either by the
period matrix or three of the six branch points. We have
\begin{eqnarray}
{A}_{II} & = &  c_{II}\, K(k_i,\epsilon_i) \, \int { \prod_{i=1}^6
{\rm d}^2 a_i/{\rm d} V_{pr} \over T^5  \,
\prod_{i<j} |a_i - a_j|^2 } \,  \nonumber \\
& & \times \prod_{i=1}^4 {  { \rm d}^2 z_i \over |y(z_i)|^2 } \,
\prod_{i<j} {\rm exp}\left[ - k_i\cdot k_j \ G(z_i,z_j) \right]
\nonumber \\
&  & \times | s (z_1z_2 + z_3 z_4) + t(z_1z_4+ z_2 z_3) + u(z_1z_3
+ z_2 z_4)|^2 , \label{eqr95}
\end{eqnarray}
where ${\rm d} V_{pr} = { {\rm d}^2 a_i {\rm d}^2 a_j {\rm d}^2
a_k \over |a_{ij}a_{jk}a_{ki}|^2}$ is a projective invariant
measure and $c_{II}$ is a constant which should be determined by
factorization or unitarity (of the $S$-matrix).

An immediate application of the above result is to study the
perturbative correction to the $R^4$ term at two loops. By taking
the limit of $k_i \to 0$, one sees from eq.~(\ref{eq111}) that the
integrand ${\cal A}_{II}$ is zero identically in moduli space
apart from the kinematic factor $K(k_i,\epsilon_i)$. So we
conclude that the perturbative correction to the $R^2$ term is
zero, confirming the explicit calculation of Iengo \cite{Iengo}.
This result is also in agreement with the non-perturbative
conjecture of \cite{GreenGutperle} (see also \cite{Green2,
GreenSethi, IengoZhu3}). Our new result also explicitly verifies
the claim given in the the Appendix B of \cite{Stieberger}.

The finiteness of the amplitude can also be checked by following
the detailed discussions given in \cite{IengoZhu3}.

\section*{Appendix A: The $\langle \partial X(z) \partial
X(w) \rangle$ correlators}

For the $X(z,\bar z)$ correlators, D'Hoker and Phong in
\cite{DHokerPhong3} use the following expression in terms of the
prime form:
\begin{equation}
\langle \partial_z X^\mu (z)
\partial_w X^\nu (w) \rangle   =   -\delta^{\mu\nu}\partial_z
\partial_w \ln E(z,w) \, ,
\end{equation}
in their effective rules for chiral splitting. In this paper we
use the following result given in \cite{Knizhnik, Zhu}:
\begin{eqnarray}
& & \langle \partial X^\mu(z) \partial X^\nu (w) = -
\delta^{\mu\nu} \, \left( {1\over 4(z-w)^2} + {1\over 4 T} \,
{\partial \over \partial w} \left[ {y(w)\over y(z)}\, {1\over
z-w}\right. \right. \nonumber \\
& &  \times  \int \left. \left.  { (z-z_1)(z-z_2)\over
(w-z_1)(w-z_2) }\, \left| z_1 - z_2 \over y(z_1)y(z_2) \right|^2
{\rm d}^2 z_1 {\rm d}^2 z_2\right]\right) + (z \leftrightarrow w).
\label{eq27}
\end{eqnarray}
This correlator satisfies the following relation:
\begin{equation}
\int \bar\Omega_i(\bar z) \langle \partial X^\mu(z) \partial
X^\nu(w) \rangle \, {\rm d}^2 z = 0.
\end{equation}
These two formulas are actually related by using eq.~(\ref{eq110})
by differentiating with respect to $z$ and $w$. We have
\begin{eqnarray}
\partial_z\partial_w \ln E(z,w)  & = &  \pi \, \omega(z) \cdot
 ({\rm Im} \tau)^{-1} \cdot  \omega (  w) \nonumber \\
 & & \hskip -2cm  + \left\{ \left( {1\over 4(z-w)^2} + {1\over 4 T} \,
{\partial \over \partial w} \left[ {y(w)\over y(z)}\, {1\over
z-w}\right. \right. \right. \nonumber \\
& &  \hskip -3.5cm \left. \times  \int \left. \left.  {
(z-z_1)(z-z_2)\over (w-z_1)(w-z_2) }\, \left| z_1 - z_2 \over
y(z_1)y(z_2) \right|^2 {\rm d}^2 z_1 {\rm d}^2 z_2\right]\right) +
(z \leftrightarrow w) \right\} \, . \label{eq333}
\end{eqnarray}

By using eq.~(\ref{eq27}), one can also derive the correlators for
$ \sum_i \langle \partial X(z) k_i \cdot X(w_i)\rangle$ with
$\sum_i \alpha_i = 0 $. It is
\begin{eqnarray}
  \sum_i  \langle \partial X^\mu(z) k_i\cdot X (w_i) \rangle & =
 & -\sum_i k_i^\mu \left\{ { 1\over 2(z-w_i) } + { 1\over 2 T} \int {
y(w_i)\over y(z) } \, { 1\over z - w_i} \right. \nonumber \\
& &  \hskip -2cm \times \left.  { (z-z_1)(z-z_2) \over
(w_i-z_1)(w_i-z_2) } \, \left| z_1-z_2 \over y(z_1)y(z_2)
\right|^2 {\rm d}^2z_1 {\rm d}^2 z_2 \right\} . \label{eq115}
\end{eqnarray}
The above correlator also satisfies
\begin{equation}
\int  \bar\Omega_i(\bar z) \sum_i \langle \partial X^\mu(z) k_i
\cdot X  (w_i)\rangle    \, {\rm d}^2 z = 0.
\end{equation}
By differentiating with respect to $\bar w_i$ in
eq.~(\ref{eq115}), we have
\begin{equation}
\langle \partial X^\mu(z)  \bar{\partial} X^\nu  (\bar w)\rangle =
- \delta^{\mu\nu}\left[ \pi \delta^{(2)}(z-w) - { \pi \over 2} \,
\omega(z) \cdot ({\rm Im} \tau )^{-1} \cdot \bar{\omega}(\bar w)
\right] ,
\end{equation}
where
\begin{eqnarray}
Q & \equiv &   \omega(z) \cdot ({\rm Im} \tau )^{-1} \cdot
\bar{\omega}(\bar w)  \nonumber \\
& = & { 2 \over T} \, { 1\over y(z) \bar y(\bar w)} \, \int {
(z-u)(\bar w - \bar u) \over |y(u)|^2 } \, {\rm d}^2 u ,
\end{eqnarray}
and
\begin{equation}
T = \int {\rm d}^2 z_1 {\rm d}^2 z_2 \, \left| z_1 - z_2 \over
y(z_1) y(z_2) \right|^2    = 2 |{\rm det}\, K |^2 \, {\rm det}
\,{\rm Im }\tau   .
\end{equation}

The $T_1$ used in eq.~(\ref{eq45}) can also be obtained from
eq.~(\ref{eq27}) explicitly. We have
\begin{eqnarray}
T_1(w) & = & {1\over32}\left(\sum_{i=1}^6{1\over w-a_i}\right)^2
- {1\over 16}\sum_{i=1}^6{1\over(w-a_i)^2} \nonumber \\
& & +{1\over 8T}\int{\rm d}^2z_1{\rm d}^2z_2 \left|{z_1-z_2 \over
y(z_1)y(z_2)}\right|^2 \left[ {2\over (w-z_1)(w-z_2)} \right.
\nonumber\\
& & + 2 \sum_{i=1}^2 {1\over (w-z_i)^2}  - \left.  \left({1\over
w-z_1}+ {1\over w-z_2}\right)\sum_{i=1}^6{1\over w-a_i}\right] .
\end{eqnarray}

The other formula used in the main text is
\begin{equation}
\partial_{p_a}\partial_{q_1} \ln E(p_a,q_1) + \partial_{p_a}
\partial_{q_2} \ln E(p_a,q_2) = { 1\over (q- p_a)^2} ,
\label{eq29}
\end{equation}
which can  be derived easily from eq.~(\ref{eq333}) by noting
$y(q_2) = - y(q_1)$ and  $\omega_i(q_2) = - \omega_i(q_1)$.

\section*{Appendix B: The proof of eq.~(\ref{eq999})}

The proof of the  summation formulas is quite simple. The strategy
is the same as the proof for other ``vanishing identities" given
in \cite{AllZhu2} which was first used in \cite{IengoZhu1,
IengoZhu2, Zhu}. We first change $S_m(x)$ into a polynomial by
using the following  M\"obius transformation:
\begin{equation}
a_i = x - { 1\over \tilde a_i} , \qquad \hbox{or} \qquad \tilde
a_i = { 1\over x - a_i}, \quad \tilde z_i = { 1\over x -z_i}.
\end{equation}
Noting that $y(z) = \prod_{i}^6(z-a_i) =
\prod_{i=1}^3(z-A_i)(z-B_i)$, we can write $u(z)$ and ${1\over
u(z)} $ as follows:
\begin{eqnarray}
u(z) & = & {\prod_{i=1}^3 (z-A_i) \over y(z)}, \\
{1\over u(z)} & = & {\prod_{i=1}^3 (z-B_i) \over y(z)}.
\end{eqnarray}
By using these expressions we see that the left hand side of eq.
(\ref{eq999}) is a homogeneous polynomial in $\tilde a_i$ and
$\tilde z_k$ (the M\"obius transformed $a_i$ and $z_k$) times $(
 \prod_{i=1}^4 \tilde y(\tilde z_k))^{-1}$, i.e.
\begin{eqnarray}
P_{n,m} & = &  \sum_\delta \eta_\delta Q_\delta \left\{ {u(z_1)
u(z_2) \over u(z_3)u(z_4)} - (-1)^n {u(z_1)u(z_2)\over
u(z_3)u(z_4) }
\right\} (S_m(x))^n \nonumber  \\
&   =  &  { y^4(x) \over \prod_{i=1}^4 \tilde y(\tilde z_i) }
\sum_\delta  \eta_\delta \tilde Q_\delta \left[
\prod_{i=1}^3\prod_{j=1}^2 \prod_{l=3}^4(\tilde z_k-\tilde
A_i)(\tilde z_l-\tilde B_i) -
(\tilde A\leftrightarrow \tilde B) \right] \nonumber \\
&  & \qquad    \times \left( \sum_{i=1}^3[ (\tilde A_i)^m -
(\tilde B_i)^m] \right)^n . \label{eq9999}
\end{eqnarray}
This polynomial is modular invariant and it is also vanishing when
$\tilde z_1 = \tilde z_{3,4}$ and $\tilde z_2 = \tilde z_{3,4}$.
So it is proportional to $P(\tilde a) \prod_{i=1}^2\prod_{j=3}^4
(\tilde z_i-\tilde z_j)$. The rest factor can only be obtained by
explicit computation (most easily by computer). Then we have
\begin{equation}
P_{n,m} = { y^4(x) \over \prod_{i=1}^4 \tilde y(\tilde z_i) }
\times { 2 P(\tilde a) \prod_{i=1}^2\prod_{j=3}^4 (\tilde
z_i-\tilde z_j)} \, \times C_{n,m}. \label{qq99}
\end{equation}
By an inverse M\"obius transformation we have \begin{eqnarray}
\tilde y(\tilde z) & = & \prod_{i=1}^6(\tilde z- \tilde a_i)^{1/2}
= { y(z) \over (x-z)^3 y(x) }, \\
P(\tilde a) & = & {P(a) \over ( y(x))^{10} }, \end{eqnarray} etc.
By using these results in eq.~(\ref{qq99}), we obtained the
results given in eqs. (\ref{eq991})--(\ref{eq992}). This completes
the proof of eq.~(\ref{eq999}).

\section*{Appendix C: The $q$ dependence of the factor ${\cal Z}$}

In this appendix we will compute explicitly the $q$ dependence of
the factor ${\cal Z}$ defined in eq.~(\ref{eqcalx}). The relevant
part is
\begin{equation}
\int {\cal D}\beta {\cal D} \gamma\, \delta( \beta(q_1)) \,
\delta( \beta(q_2)) \, {\rm e}^{ i S_{\beta\gamma}} = ({\rm
det}'\bar{\partial}_{3/2})^{-1}\, \int {\rm d}\beta_1 {\rm
d}\beta_2 \, \delta( \beta_0(q_1))\delta_0( \beta(q_2)) ,
\end{equation}
where $\beta_1$ and $\beta_2$ are the two coefficients in the zero
mode part of the ghost field $\beta(z)$:
\begin{equation}
\beta_0(z) = \beta_1 \varphi_1^*(z) + \beta_2
\varphi^*_2(z),
\end{equation}
where $\varphi^*_{1,2}(z)$ are two holomorphic $3 \over2
$-differentials. Comparing with the previously defined two
holomorphic $3\over 2$-differential $\psi^*_{1,2}(z)$,
$\varphi^*_{1,2}(z)$ are not required to satisfy the
``normalization" condition eq.~(\ref{psinormal}).

Choosing an arbitrary 1-differential $\omega_x(z)$ which has two
zeroes at $z=x_1 = x+$ and $z=x_2 = x-$, the $\varphi^*_{1,2}(z)$
can be constructed as follows:
\begin{equation}
\varphi_\alpha^*(z) = \omega_x(z)\, S(z,x_\alpha).
\end{equation}
By using these holomorphic differentials, we have
\begin{eqnarray}
& & \int {\rm d}\beta_1 {\rm d}\beta_2 \, \delta(
\beta_0(q_1))\delta( \beta_0(q_2)) \nonumber \\
& & \qquad = \int {\rm d}\beta_1 {\rm d}\beta_2 \,
\prod_{\alpha=1}^2 \delta( \beta_1 \omega_x(q_\alpha)
S(q_\alpha,x_1) + \beta_2 \omega_x(q_\alpha) S(q_\alpha,
x_2) ) \nonumber \\
& & \qquad = |{\rm det}(\omega_x(q_\alpha) S(q_\alpha,x_\beta) )
|^{-1} = \left[ q-x \over \omega_x(q)\right]^{2} ,
\end{eqnarray}
by using the explicit expression of the Szeg\"o kernel and noting
that $\omega_x(q_2) = -  \omega_x(q_1)$ and $u(q_2) = - u(q_1)$,
etc.  In hyperelliptic language, the 1-differential $\omega_x(z)$
is given as follows:
\begin{equation}
\omega_x(z) = { c(z- x)\over y(z)},
\end{equation}
where $c$ is a possibly moduli dependent constant. By using this
result we have:
\begin{equation}
 \int {\cal D}\beta {\cal D} \gamma\, \delta( \beta(q_1)) \,
\delta( \beta(q_2)) \, {\rm e}^{ i S_{\beta\gamma}}  = \langle
\delta( \beta(q_1)) \, \delta( \beta(q_2))  \rangle = { y^2(q)
\over c^2 } \,\times ({\rm det}'\bar{\partial}_{3/2})^{-1}.
\end{equation}
The $q$ dependent factor $y^2(q)$ exactly cancels the factor
$y^2(q)$ in ${\cal A}$.  The complete expression for ${\cal Z}$
is:
\begin{equation}
{\cal Z} = { y^2(q)  \, \prod_{i<j} A_{ij} B_{ij}
 \over c^2 ({\rm det} K)^5\, \prod_{i<j} (a_i - a_j) }
 = { y^2(q)  \,   Q_\delta \over c^2 ({\rm det} K)^5\,  P(a) }  ,
\end{equation}
by using the results of \cite{IengoZhu2} (for previous works see
\cite{Dixon, Zamolodchikov, Radul, Knizhnik2, Montano, Gava}).

Finally we  have a variational formula from \cite{Morozov2}:
\begin{equation}
{\partial \tau_{ij} \over \partial a_n} = { i \pi \over 2 } \,
\hat\omega_i(a_n) \hat\omega_j(a_n) \, , \label{tautoa}
\end{equation}
where $\omega (a_n)$ is defined as follows:
\begin{eqnarray}
\omega (z) {\rm d} z & = & ( \hat\omega (z_0) +
\hat\omega'(z_0)(z-z_0) +
\cdots ) {\rm d} z \nonumber \\
& = & 2 u \omega(u^2 + a_n) \, {\rm d} u,
\\
\omega(a_n) & = &   \lim_{u \to 0 } 2 u \omega( u^2 + a_n) .
\end{eqnarray}
Here we have used the uniformization  coordinate $u$ instead of
$z$ at the branch point $a_n$.

By using eq.~(\ref{tautoa}), we have:
\begin{equation}
{\partial ( \tau_{ij} )\over \partial(a_1 a_2 a_3) } = i \left(
\pi \over 2 \right)^3 \, { a_{45}a_{46} a_{56} \over ({\rm det}
K)^3 P(a) } ,
\end{equation}
which can be used to transform from period matrix parametrization
to the branch points parametrization of the moduli space.

\section*{Acknowledgments}

Chuan-Jie Zhu would like to thank Roberto Iengo for reading the
paper and comments. He would also like to thank E. D'Hoker and D.
Phong for discussions and Jian-Xin Lu and the hospitality at the
Interdisciplinary Center for Theoretical Study, Physics,
University of Science and Technology of China.

\end{document}